\documentclass[10pt,letterpaper]{article}
\usepackage{opex3}
\usepackage[separate-uncertainty=true]{siunitx}

\begin{document}

\title{Generating polarization-entangled photon pairs using cross-spliced birefringent fibers}

\author{Evan Meyer-Scott,$^{1*}$ Vincent Roy,$^2$ Jean-Philippe Bourgoin,$^1$ Brendon L. Higgins,$^1$ Lynden K. Shalm,$^1$ and Thomas Jennewein$^1$}

\address{$^1$Institute for Quantum Computing and Department of Physics and Astronomy, University of Waterloo, Waterloo, Ontario, N2L 3G1, Canada\\
$^2$Institut National d'Optique, 2740 Rue Einstein, Quebec City, Quebec, Canada}
\email{*emeyersc@uwaterloo.ca}


\begin{abstract}We demonstrate a novel polarization-entangled photon-pair source based on standard birefringent polarization-maintaining optical fiber. The source consists of two stretches of fiber spliced together with perpendicular polarization axes, and has the potential to be fully fiber-based, with all bulk optics replaced with in-fiber equivalents. By modelling the temporal walk-off in the fibers, we implement compensation necessary for the photon creation processes in the two stretches of fiber to be indistinguishable. Our source subsequently produces a high quality entangled state having \SI{92.2 \pm 0.2}{\percent} fidelity with a maximally entangled Bell state.\end{abstract}



\section{Introduction}


Many tasks in quantum information, such as loophole-free tests of Bell's inequality~\cite{PhysRevLett.81.5039}, long-distance quantum cryptography~\cite{1367-2630-11-8-085002}, quantum metrology~\cite{Giovannetti19112004}, and integrated quantum information processing~\cite{Martin-Lopez2012Experime}, require the generation and detection of entangled photon pairs with high efficiency. While recent advances using superconductors have led to detectors with efficiencies over \SI{90}{\percent}, these detectors require the photons to be fiber-coupled or integrated with a wave\-guide~\cite{2012arXiv1209.5774M,2012arXiv1209.5721L}.  Simple low-loss and integrable sources of entangled photons remain a challenge. Sources based on spontaneous parametric down-conversion in bulk crystals, where a pump photon fissions into two daughter photons, are challenging to couple into single-mode fiber. Much work has recently focused on using the process of four-wave mixing to create entangled photon pairs directly in fiber~\cite{PhysRevA.73.052301}. Current sources either require interferometric stability, have excess noise, or suffer from losses in bulk optical components. Here we present a fiber source of entangled photons that is simple, stable, integrable with all-fiber components, and is well-suited for carrying out a wide variety of quantum information processing tasks.

\section{Entangled photon generation in birefringent optical fibers}

To generate entanglement, two processes distinct in some degree of freedom must interfere. One of the most successful sources of entangled photon pairs uses two nonlinear crystals back-to-back with their optic axes rotated \SI{90}{\degree} with respect to one another~\cite{PhysRevA.60.R773}. The interference between the possibility of down-conversion in the first or second crystal, upon which the polarization of the photon pairs depends, leads to the creation of an entangled polarization state between the two photons. Inspired by this ``sandwiched'' cross-crystal source, we have devised a cross-spliced source~\cite{note1} using four-wave mixing in birefringent polarization-maintaining (PM) fiber. In the fiber, two pump photons can spontaneously interact to produce a pair of photons (labeled signal and idler). The signal and idler photons possess the same polarization, perpendicular to that of the pump photons. To create entanglement, we take two PM fibers, rotate them \SI{90}{\degree} with respect to one another, and splice them together. Pump photons polarized at \SI{45}{\degree} with respect to the fiber axes will be equally likely to generate photon pairs in either fiber section, leading to the creation of the polarization entangled state $|\phi \rangle = ( |HH\rangle + e^{i\phi}|VV\rangle )/\sqrt{2}$, where $H$ and $V$ are horizontal and vertical polarizations respectively. This single-path geometry permits direct fiber splicing to the output, is interferometrically stable, and allows the source to be compact. In comparison, other schemes for generating entangled photon pairs in fiber require polarization stabilization~\cite{1367-2630-9-8-289,Medic:10}, suffer from losses and imperfections in the optical components needed to interfere the polarizations of the photons~\cite{Liang:06,Hall:09}, or produce photons very close to the pump leading to unwanted Raman noise~\cite{Zhou:12}. Approaches based on a fiber Sagnac loop are promising~\cite{PhysRevA.76.043836} but prone to imperfect interference on the beam-splitter and temporal walk-off due to beam-splitter birefringence~\cite{Fang:12}.

To create the signal and idler photons through four-wave mixing, energy and momentum must be conserved. Energy conservation is satisfied when the sum of the frequencies of the two pump photons equals the sum of the signal and idler frequencies. In our scheme, momentum conservation, or phase-matching, is engineered using the birefringence in our PM fibers~\cite{PhysRevA.79.023840}. PM fibers can support vector phase-matching wherein the pump is polarized on the slow axis of the fiber and the outgoing signal and idler photons on the fast axis~\cite{note2, Smith:09,Halder:09,NarrowPCF,PhysRevLett.102.123603}. This form of phase-matching is intrinsically narrowband and has the signal and idler photons located spectrally far from the pump ($\sim $\SI{100}{\nano\meter} or $\sim $\SI{50}{\tera\hertz})~\cite{PhysRevA.79.023840}. This reduces the noise from Raman scattering, which extends about \SI{15}{\tera\hertz} on either side of the pump but has a long tail on the low-frequency side~\cite{PhysRevA.75.023803} that can dramatically degrade the heralding efficiency (probability of detecting signal (idler) photon given an idler (signal) detection) and entanglement fidelity. An additional benefit is that two-photon states free of frequency correlations can be created in PM fibers by controlling only the fiber length and pump bandwidth~\cite{Smith:09}, or by pumping with two different wavelengths~\cite{FangLorenz:12}. These spectrally pure photon states are critical for multi-photon interference experiments and for quantum computing protocols, as otherwise tight inefficient spectral filtering of the photon pairs is needed. Using a similar phase-matching scheme, heralded (unentangled) photons with a purity of up to \SI{84}{\percent}~\cite{PhysRevA.83.031806} have been demonstrated.

\section{Phase compensation}

In our cross-spliced source, photons produced in the first fiber will be partially distinguishable from those produced in second. This distinguishability is due to the poor temporal overlap caused by additional dispersion that photons generated in the first fiber experience in the second. It is convenient to reframe this temporal problem as a spectral one: the phase $\phi$ in the maximally entangled state $|\phi \rangle = ( |HH\rangle + e^{i\phi}|VV\rangle )/\sqrt{2}$ becomes wavelength dependent due to the different effective fiber lengths seen by different spectral components~\cite{Trojek2007Efficien,trojek:211103}. This leads to a mixture over the pump and output signal bandwidths~\cite{note3} of the form
\begin{equation}
\rho = \int\int |\phi\rangle\langle\phi | p_\text{s}(\lambda_\text{s}) p_\text{p}(\lambda_\text{p}) \, \mathrm{d}\lambda_\text{s} \mathrm{d}\lambda_\text{p},
\label{eqn.mixture}
\end{equation}
where $p_\text{s}(\lambda_\text{s})$ and $p_\text{p}(\lambda_\text{p})$ describe the spectra of the pump and signal respectively, and $\lambda_\text{p/s/i}$ is the wavelength of the pump/signal/idler.
Note that Eq.~(\ref{eqn.mixture}) is not describing the spectral components of the output photon state but, rather, the effective two-qubit state given the spectral dependence of the phase $\phi$. To produce high quality entanglement we must find and compensate for this spectral dependence.

We assume the two fiber segments have the same length $L$ and that the first fiber has its slow axis vertical such that $|HH\rangle$ photon pairs are produced first. These photons acquire an extra phase
\begin{equation}
\phi_1(\lambda_\text{s},\lambda_\text{p}) = \frac{2\pi L}{\lambda_\text{s}}\left[n(\lambda_\text{s})+B\right] + \frac{2\pi L}{\lambda_\text{i}}\left[n(\lambda_\text{i})+B\right]
\end{equation}
in the second length of fiber, where (from energy conservation) $\lambda_\text{i} = \lambda_\text{s} \lambda_\text{p} / (2\lambda_\text{s} - \lambda_\text{p})$, $n(\lambda)$ is the propagation constant on the fast axis in the fiber, and $B$ is the birefringence between the fiber's fast and slow axes. Additionally, the $|VV\rangle$ term acquires a phase 
\begin{equation}
\phi_2(\lambda_\text{p})=2\frac{2\pi L}{\lambda_\text{p}}\left[n(\lambda_\text{p})\right] \label{eqn.vvphase}
\end{equation}
in the first stretch of fiber via the pump. This phase is double that of its counterpart for $\chi^{(2)}$ sources~\cite{Trojek2007Efficien} because here two pump photons combine to make the $|VV\rangle$ photon pairs. There are no additional $B$ terms in Eq.~(\ref{eqn.vvphase}) because the pump that will produce $|VV\rangle$ photons in the second stretch of fiber is polarized on the fast axis in the first stretch.

Finally, there is a nonlinear contribution to the $|VV\rangle$ phase term from self- and cross-phase modulation of the pump,
\begin{equation}
\phi_{2,\text{NL}}(\lambda_p)=\left[1+(2/3)\right]\gamma PL, \label{eqn.nlphase}
\end{equation}
where $\gamma$ is the fiber's nonlinear parameter and $P$ is the pulse peak power. This contributes only a small offset in our pumping regime, but could become important at very short pulse length, high intensity, or for frequency-chirped pulses.

\begin{figure}[htbp]
\centering
 \includegraphics[width=.7\linewidth]{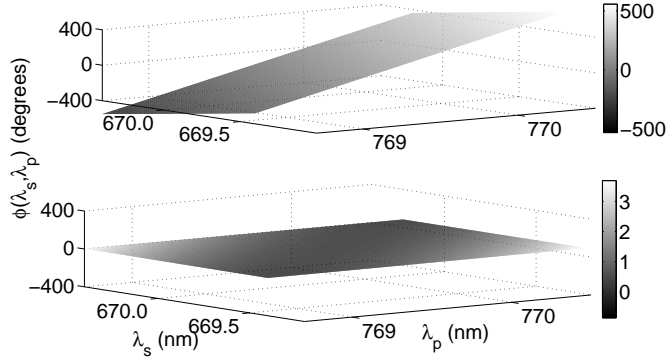}
 \caption{Theoretical phase deviation around the mean phase, before (top) and after (bottom) compensation, for various signal and pump wavelengths. The strong phase change over pump and signal bandwidths in the uncompensated case leads to a highly mixed state, while the nearly flat map after compensation leads to a nearly pure entangled state.\label{fig.phase}}
 \end{figure}

 The total phase is then 
\begin{equation}
\phi(\lambda_\text{s},\lambda_\text{p})=\phi_2+\phi_{2,\text{NL}}-\phi_1.
\label{eqn.phi}
\end{equation}
This phase is plotted in Fig.~\ref{fig.phase} and displays a variation of \SI{800}{\degree} over the pump and signal bandwidths. To compensate this variation, birefringent crystals of appropriate length are introduced into the signal and idler output arms, adding two controllable phases to Eq.~(\ref{eqn.phi}). After optimizing, the compensators flatten the phase profile in Fig.~\ref{fig.phase} (the edges of the phase profile are about \SI{5}{\degree} above zero). While we use quartz wedges with variable thicknesses, a production source could use precisely cut lengths of most any birefringent material. With $L = \SI{13}{\centi\meter}$, the optimal total length of quartz
is calculated to be \SI{67.3}{\milli\meter} with slow axis vertical on the signal arm (here implemented by rotating the polarization of the signal photons by \SI{90}{\degree}), and \SI{47.6}{\milli\meter} with slow axis horizontal on the idler arm. This flattens the phase map sufficiently that no drop in entanglement fidelity is expected due to phase/temporal distinguishability.

\section{Experimental demonstration}

 \begin{figure}[htbp]
\centering
 \includegraphics[width=.8\linewidth]{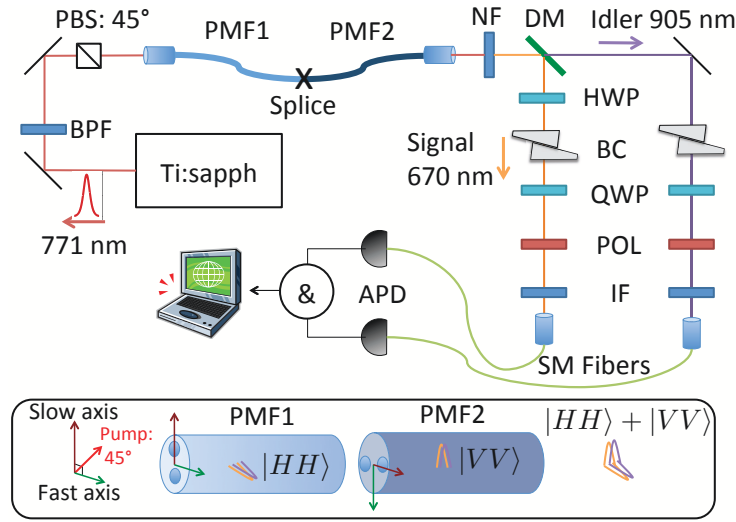}
 \caption{Experimental apparatus for the cross-spliced entanglement source. Pulses generated by the mode-locked Ti:sapphire laser pass though a bandpass filter (BPF) centred at \SI{771}{\nano\meter} wavelength, then through a polarizing beam-splitter (PBS) oriented at \SI{45}{\degree}. This light is coupled into the cross-spliced fibers (PMF1 and PMF2) where entangled photon pair generation occurs via four-wave mixing. The pump is removed with a notch filter (NF) and the signal and idler photons are separated with a dichroic mirror (DM). The signal photon passes through a half-wave plate (HWP) and both photons are phase/time compensated with tuneable birefringent compensators (BC). The polarization correlations of the photons are analyzed with quarter-wave plates (QWP) and polarizers (POL) before photons pass through interference filters (IF) and are coupled into single-mode (SM) fibers. The photons are detected by silicon avalanche photodiodes (APD) and their detection and coincidence counts are registered by a timetagging module (\&), then recorded by a computer. Inset: illustration of the polarization axes of crossed-fibers and photon generation.\label{fig.exp}} 
 \end{figure}

The experimental apparatus is shown in Fig.~\ref{fig.exp}. The cross-splice process was carried out using an arc fusion splicer which provides automated alignment of the polarization axes of PM fibers (FSM-45PM, Fujikura Ltd.). The source is pumped with a mode-locked Ti:sapphire laser at \SI{76}{\mega\hertz} repetition rate,  \SI{3}{\pico\second} pulse length, and full width at half maximum bandwidth of \SI{0.3}{\nano\meter}, with polarization set to \SI{45}{\degree}. It is important to include a bandpass filter on the pump beam to remove spontaneous emission from the Ti:sapphire crystal, which would pollute the photon pair signal. After photon generation in the two cross-spliced \SI{13}{\centi\meter} sections of PM630-HP fiber, the pump light is removed with a notch filter providing \SI{60}{\decibel} isolation. The half-wave plate in the signal arm flips the polarization of the signal photons by \SI{90}{\degree}, so that they are correctly compensated by the birefringent quartz crystals. 

\begin{figure}[htbp]
 \centering
 \includegraphics[width=\linewidth]{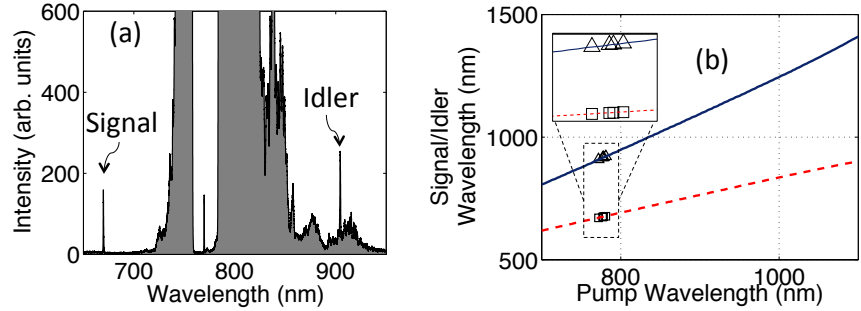}
 \caption{(a) Measured spectrum from our cross-spliced fiber source, showing narrowband signal and idler modes well away from the main Raman contamination, which extends beyond the vertical range shown. Some residual pump in the centre of the notch filter is also visible. (b) Theoretical phase-matching curves for signal (dashed red) and idler (solid blue)  with measured points indicated by squares and triangles, respectively. A wide range of signal and idler wavelengths are available in PM fiber by tuning the pump wavelength. Other vector and scalar phase-matching conditions exist (see e.g.\ Refs.~\cite{PhysRevA.79.023840,Halder:09,NarrowPCF}) but are not relevant here.
 \label{fig.spec}}
 \end{figure}

To characterize the output from the source, we collected signal and idler photons and sent them to an optical spectrometer and cooled CCD camera (SpectraPro 2750 and Spec-10-LN model 7508, Princeton Instruments). A sample spectrum is shown in Fig.~\ref{fig.spec}(a) where the signal and idler peaks are clearly identifiable, and far enough from the pump to avoid the bulk of the Raman noise. The phase-matching curve in Fig.~\ref{fig.spec}(b) demonstrates the wide tunability of our source, with the idler photon ranging from the visible to telecom bands as the pump wavelength is changed. In this work, the \SI{771}{\nano\meter} pump generates signal photons at \SI{670}{\nano\meter} and idler photons at \SI{905}{\nano\meter}, with full width at half maximum bandwidths of 
\SI{0.23}{\nano\meter} and \SI{0.61}{\nano\meter} respectively. 

To count photon pairs, signal and idler photons exiting the source are split on a dichroic mirror, analyzed by a quarter-wave plate and polarizer, and filtered with an interference filter centered on \SI{670}{\nano\meter} for the signal and \SI{905}{\nano\meter} for the idler, each providing an additional \SI{60}{\decibel} isolation from the pump. Photons are then coupled into SM780 single-mode fiber and sent to SPCM-AQ4C single photon detectors (Perkin-Elmer). By retuning the Ti:sapphire pump laser to the signal and idler wavelengths, we saw a total coupling efficiency from the PM fiber to the output of the single-mode fiber of \SI{60}{\percent} and \SI{69}{\percent}, respectively, including the notch filter, dichroic mirror, and birefringent compensators.

\section{Performance of the cross-spliced source}

As the number of double-pair emissions is negligible for our pump powers, we can determine the total heralding efficiency of the signal and idler photons. In \SI{30}{\second} with \SI{30}{\milli\watt} pump power, we collected \num{488350} total and \num{146901} background signal counts; \num{1657630} total and \num{1435459} background idler counts; and \num{53256} total and \num{55} background coincidence counts. After subtracting background counts (due to Raman scattering, dark counts, and stray light) these results give heralding efficiencies of \SI{24}{\percent} for the signal and \SI{16}{\percent} for the idler, including detectors. This efficiency is limited mainly by our avalanche photodiodes, which have limited detection efficiency especially at the idler wavelength. Additional losses are not intrinsic to the source, but are rather due transmission through bulk optics necessary in this first demonstration.

By measuring the heralding efficiency of $|HH\rangle$ and $|VV\rangle$ photon pairs separately, we can infer the transmission of the splice to be at least \SI{93}{\percent} at the signal wavelength and at least \SI{96}{\percent} at the idler wavelength.

\begin{figure}[htbp]
 \centering
 \includegraphics[width=\linewidth]{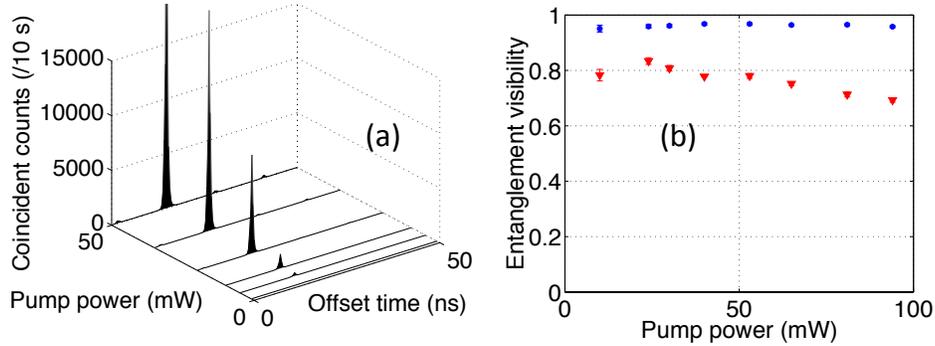}  
  \caption{(a) Measured coincidence timing histograms, showing excellent timing correlation of signal and idler photons and negligible background. The \SI{50}{\milli\watt} peak extends beyond the top of the graph, which has a shortened vertical scale in order to see accidental coincidences that occur regularly at the pump repetition period. (b) Dependence of entanglement visibility on average pump power, for rectilinear basis (blue circles) and diagonal basis (red triangles). Total visibility reaches a maximum around \SI{30}{\milli\watt} pump power. Error bars are comparable to symbol size. \label{fig.pol}}
 \end{figure}

To show that the source produces a viable entangled state, we checked both the timing and polarization correlations between photons. The timing histograms in Fig.~\ref{fig.pol}(a) quantify the arrival time of the idler conditioned on a signal photon detection, and clearly show a strong peak of coincident detections. At the highest power level it is possible to see accidental coincidences \SI{13}{\nano\second} (the laser repetition period) on either side of the main peak, but their relative strength is so low as to affect the entangled state negligibly. The coincidences-to-accidentals ratio for \SI{50}{\milli\watt} average pump power is 110, based on a \SI{1}{\nano\second} coincidence timing window, and increases to 260 at \SI{10}{\milli\watt}.

The entanglement visibility in Fig.~\ref{fig.pol}(b) is high for all power settings in the rectilinear basis (i.e.\ double-pair emissions remain negligible), but shows a degradation at high pump power in the diagonal basis. At high pump powers, we saw a noticeable broadening of the pump and therefore signal and idler spectra, due to self-phase modulation of the pump. This has the effect of distinguishing photons created in the first fiber segment from those created in second, since the pump is broader by the time it reaches second fiber, decreasing visibility in the diagonal basis. At low pump powers, the visibility also degrades due to noise from spontaneous Raman scattering, which is linear in pump power (compared to the quadratic response of four-wave mixing). Therefore, a pumping strength optimal for entanglement visibility emerges, which in our case is around \SI{30}{\milli\watt}.

\begin{figure}[htbp]
 \centering
 \includegraphics[width=.8\linewidth]{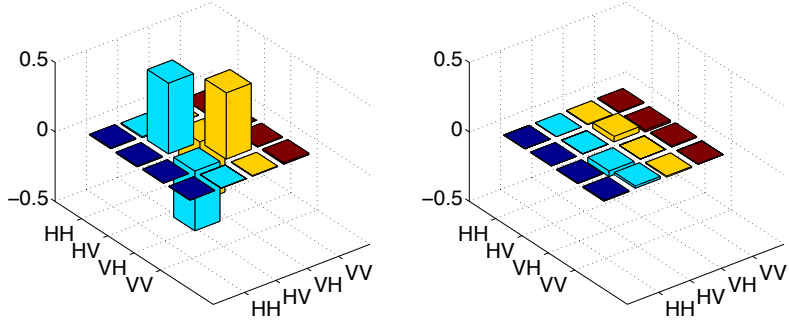}   \caption{Real (left) and imaginary (right) parts of the two-qubit tomographically reconstructed density matrix of the photon pair produced by the source, showing fidelity with the maximally entangled state singlet state $|\Psi^-\rangle$ of \num{0.922 \pm 0.002}. (The source produces this state instead of $|\Phi^-\rangle$ due to the polarization flip of the signal photon.)\label{fig.tomo}}
 \end{figure}

Pumping at \SI{33}{\milli\watt}, with a pair production rate in the fiber inferred to be \SI{45000}{pairs/\second}, we took tomographic data to reconstruct the two-qubit density matrix of our entangled state, shown in Fig.~\ref{fig.tomo}. Without any background subtraction or correction, the fidelity with the maximally entangled singlet Bell state $|\Psi^-\rangle = (|HV\rangle - |VH\rangle)/\sqrt{2}$ is \num{0.922 \pm 0.002} and the tangle is \num{0.721 \pm 0.008}.

As an important point for integrability and coupling efficiency, the bulk pump laser, filters, dichroic mirror, and birefringent compensating optics could be replaced respectively by a fiber laser~\cite{Limpert2006High-pow}, fiber Bragg gratings~\cite{Vengsarkar1996Long-per}, wavelength-division multiplexing, and birefringent fiber. These technologies have been heavily developed for the telecom band around \SI{1500}{\nano\meter} and, as seen in Fig.~\ref{fig.spec}(b), the idler photon can be tuned to this regime, while development is ongoing for these fiber technologies in the visible spectrum. Replacing our bulk optics with in-fiber equivalents would greatly decrease reflection and coupling losses.
 
\section{Conclusion}
We have demonstrated a simple source of entangled photon pairs based on a single path of standard optical fiber. Splicing two stretches of PM fiber with rotated polarization axes allows high quality entanglement without problems of spatial mode overlap or stability. The source demonstrates narrowband photons (${<}\SI{1}{\nano\meter}$), the possibility of wide tunability of output wavelengths, ${>}\SI{60}{\percent}$ optical coupling to optical fiber networks, and fidelity with a maximally entangled state of \num{0.922}. We experimentally verified that the quality of a regular fiber splice is sufficient for connecting independent four-wave mixing processes without degrading the entangled state.  Thus we envisage future experiments with multiple fiber links to compensate dispersive broadening or walk-off, or a fiber version of a superlattice for spectral engineering~\cite{PhysRevLett.97.223602}. With the goal of improving coupling and heralding efficiencies, we plan to study the reduction of Raman noise through optimization of spectral filtering, fiber cooling, and novel pumping schemes~\cite{Fan:05}. The simplicity of our new photon source architecture is expected to drive use in upcoming experiments requiring integrable sources and narrowband, tuneable photons, as well as spur thought into the possibilities afforded by direct splicing of fibers for photon sources. 

\section*{Acknowledgments}
We thank Pascal Deladurantaye and Piotr Kolenderski for helpful discussions. This work was supported by the Natural Sciences and Engineering Research Council of Canada, Canada Foundation for Innovation, Canadian Institute for Advanced Research, Canadian Space Agency, Ontario's Early Researcher Awards, and Industry Canada.

\begin{thebibliography}{10}
\bibitem{note1}
Following the trend of culinary nomenclature, we informally refer to our cross-spliced source as a ``sausage'' source.
\bibitem{note2}
Other forms of vector phase-matching with cross-polarized pump photons or cross-polarized signal/idler photons are not relevant here.
\bibitem{note3}
Due to energy conservation, the phase dependence on the idler wavelength is fully determined by the signal and pump wavelengths.
\end{thebibliography}
\end{document}